\documentclass[twocolumn]{aastex631}

\newcommand{\ha}{${\rm H\alpha}$}
\newcommand{\hb}{${\rm H\beta}$}
\newcommand{\pa}{${\rm Pa\alpha}$}
\newcommand{\nii}{\hbox{[N\,{\sc ii}]}}

\begin{document}

\title{Characterizing Dust Extinction and Spatially Resolved Paschen-$\alpha$ Emission within 97 Galaxies at $1<z<1.6$ with JWST NIRCam Slitless Spectroscopy}

\correspondingauthor{Zhaoran Liu}
\email{zhaoran.liu@astr.tohoku.ac.jp}
\author[0009-0002-8965-1303]{Zhaoran Liu}
\affiliation{Astronomical Institute, Graduate School of Science, Tohoku University, 6–3 Aoba, Sendai 980-8578, Japan}

\author[0000-0002-8512-1404]{Takahiro Morishita}
\affiliation{IPAC, California Institute of Technology, MC 314-6, 1200 E. California Boulevard, Pasadena, CA 91125, USA}

\author[0000-0002-2993-1576]{Tadayuki Kodama}
\affiliation{Astronomical Institute, Graduate School of Science, Tohoku University, 6–3 Aoba, Sendai 980-8578, Japan}



\begin{abstract}
We present results on the Paschen-$\alpha$ (\pa) emitting galaxies observed as part of the JWST FRESCO survey in the GOODS-North and GOODS-South fields. Utilizing the JWST NIRCam wide field slitless spectroscopy (WFSS), we analyze emission line fluxes, star formation rates (SFRs), and spatially resolved flux distributions of 97 \pa\ emitters at $1<z<1.6$. To assess dust extinction within our sample, we combine \pa\ fluxes with archival \ha\ data taken with the Hubble Space Telescope (HST) WFC3 G141 grism. We employ two-dimensional \pa\ and F444W mapping to trace the distributions of star formation and stellar mass, respectively. 
Our observations indicate that lower mass galaxies are almost dust free in \pa\ and exhibit smaller sizes both in star formation and underlying stellar continuum. In contrast, galaxies with stellar masses above $10^{9.5}M_\odot$ display diverse dust extinction and star formation patterns. This indicates that the structures and properties of massive galaxies evolve through different phases, which involve, e.g., star formation in massive clumps, compaction, and inside-out quenching. While average dust extinction rises with stellar mass, there is significant scatter among high-mass systems. This study demonstrates the capabilities of JWST WFSS in conducting systematic investigations of emission line galaxies and highlights the pivotal role of \pa\ in advancing our understanding of dust extinction and obscured star formation in the early universe.
\end{abstract}



\section{Introduction} \label{sec:intro}
Galaxies are vast, diverse systems evolving through complex processes over the cosmic time, with their formation and structural development significantly dependent on baryonic processes occurring within dark matter halos and interacting with molecular gas reservoirs. Understanding star formation, which is the fundamental process that shapes galaxies, is crucial for determining their structural build-up, thereby enhancing our comprehension of how galaxies are constructed and evolve.

Over the past decades, various tracers of star formation across a broad spectrum of wavelengths, from ultraviolet (UV) and optical to infrared (IR) and sub-millimeter, have been intensively studied. These multi-wavelength studies have provided invaluable insights into how galaxies form, evolve, and eventually quenched across various redshifts and environments. Previous studies have demonstrated that a large fraction of local galaxies formed most of their current stellar mass during the epoch between redshift $z=1$ and $z=3$ \citep[e.g.,][]{Madau14,behroozi19}. 
However, a prominent question remains unanswered: how exactly do galaxies assemble their stellar mass via star formation? Spatially resolved emission line mapping inside galaxies has played a crucial role in addressing this question, offering deeper insights into the mechanisms behind stellar accumulation in galaxies.

Prior to the JWST era, cutting-edge instruments with sub-kiloparsec resolving power had made it possible to resolve the star formation processes within galaxies. HST grism enabled studies of \ha\ line maps, uncovering extended star-forming H{\footnotesize II} regions. Combined with stellar mass maps, these observations helped elucidate the processes of galaxy mass assembly \citep[e.g.,][]{nelson12, Wuyts13, vulcani15, nelson16, Matharu21}. FIR and (sub)millimeter facilities such as the Atacama Large Millimeter/submillimeter Array (ALMA) expanded this view into longer wavelength, offering high-resolution insights into the dust continuum and molecular gas dynamics, revealing the compact continuum of dusty sub-mm galaxies \citep[e.g.,][]{simpson15, hodge16, barro16, tadaki17, elbaz18, Calistro18, fujimoto18, tadaki20, chen20, cheng20, ikeda22, guijarro22}. When ionized gas, dust continuum, and molecular gas observations are combined, we see a detailed picture of galactic mass assembly as a blend of extended, young star-forming disk and a centrally concentrated, extremely dusty bulge. However, interpreting \ha\ data presents challenges due to non-uniform dust extinction across different parts of a galaxy, with ongoing investigations into how this spatial variation evolves with redshifts, stellar masses, and SFRs \citep[e.g.,][]{tadaki14, nelson16_balmer,Matharu23, liu23_smacs}. Furthermore, studies aimed at resolving gas and dust components also face limitations --- they are often restricted to relatively bright sources or require significant allocation of facility's resources. Moreover, applying these observations to understand star formation relies on assumptions about dust temperature and the relationships between gas mass and star formation, which can vary not only from one galaxy to another but also within individual galaxies \citep{Shetty13, Liang19, Feldmann20}.


With the launch of the JWST \citep{Gardner06, Gardner2023}, many previous constraints have been overcome, enabling us to study star formation using a broader array of tracers and in greater detail thanks to its high-quality imaging from the optical to the MIR (0.6 -- 28.3 $\mu$m) and the spatial and spectral resolving power of its spectrograph. A significant advancement includes improved determination of dust extinction; the Near Infrared Spectrograph (NIRSpec) facilitates the study of multiple Balmer and Paschen lines, allowing for more accurate dust extinction measurements in high-$z$ galaxies \citep[e.g.,][]{Shapley23, Reddy23, Sandles23, Bunker23, morishita24wr}. 

Spatially resolving dust extinction within galaxies has also been greatly enhanced by the Near Infrared Imager and Slitless Spectrograph (NIRISS) and NIRCam Wide Field Slitless Spectrograph (WFSS). These advanced instruments can obtain spectra for all galaxies within their fields, enabling detailed observation of the nebular emission line distribution inside galaxies without pre-selection. The breakthrough extends not only to high-redshift observations but also to the study of longer-wavelength hydrogen recombination lines, which are less affected by extinction, such as the \pa\ line at 1.875 $\mu$m. With the NIRCam WFSS F444W, we can resolve \pa\ emission lines up to $z=1.7$ \citep{Neufeld24}. As an independent star formation tracer, \pa\ stands out because it closely tracks the intrinsic SFRs compared to the Balmer lines, thus providing constraints on dust extinction models and offering a more robust tracer of star formation \citep{Reddy23}.

In this work, we utilize the JWST NIRCam WFSS grism to study dust extinction and spatially resolved star formation in galaxies at $1.0<z<1.6$. Our goal is to evaluate the degree of dust extinction in our galaxy sample, assess the effects of dust on the \pa\ emission line, and employ \pa\ as a star formation tracer to study galaxy mass assembly. We detail our data reduction methods and sample selection in Sec. \ref{sec:data}. Our techniques for analyzing stellar mass, dust extinction, and size measurements are detailed in Sec. \ref{sec:analysis}. We present our results in Sec. \ref{sec:results} and compare them with the existing literature in Sec. \ref{sec:discussion}. We summarize our key conclusions in Sec. \ref{sec:summary}.

In this paper, we adopt the AB magnitude system \citep{oke83,Fukugita96}, cosmological parameters of $\Omega_m=0.3$, $\Omega_\Lambda=0.7$, $H_0=70$\,km\,s$^{-1}\,{\rm Mpc}^{-1}$, and the \citet{Chabrier03} initial mass function (IMF).

\section{DATA AND SAMPLE SELECTION} \label{sec:data}
The observational dataset utilized in this work encompasses three primary components: (1) extensive optical imaging from HST and detailed broad- and medium-band imaging obtained with the JWST/NIRCam; (2) JWST/NIRCam WFSS F444W grism; and (3) HST/WFC3 G141 grism. We detail the photometric and spectroscopic data and our sample selection as follows.

\subsection{NIRCam+HST Photometry and Photmetric Redshift}\label{sec:nircam catalog}
For source identification, we start with a catalog generated in \citet{Morishita24}. 
The work retrieved the fully processed images and spectroscopic catalogs made available by the JADES team (\citealt{Rieke23}). A photometric catalog was then constructed following the method described in \citet{Morishita24}. In our following analysis, we adopt the aperture flux from the catalog, which was corrected to represent the total fluxes of the galaxies. For further details regarding the catalog, the readers are referred to \citet{Morishita24}. 
\subsection{NIRCam WFSS Reduction}\label{sec:wfss}
In this analysis, we utilize publicly available NIRCam WFSS data from the First Reionization Epoch Spectroscopic Complete Survey (FRESCO; GO-1895; PI: P. Oesch; \citealt{Oesch23}). FRESCO is a JWST Cycle 1 program which features comprehensive deep NIRCam imaging and slitless spectroscopic observations using the F444W filter. The obtained spectra achieve a resolution of R $\sim$ 1600 and covers a wavelength range from 3.8 to 5.0 $\mu$m, which enables us to study multiple hydrogen recombination lines across a wide range of cosmic time with a spatially resolved view. 

We follow the methodology outlined in \citet{Sun23_grism}\footnote{\href{https://github.com/fengwusun/nircam\_grism/}{https://github.com/fengwusun/nircam\_grism/}} and reduce the WFSS data using a combination of the official JWST calibration pipeline \citep{Bushouse23} and several customized steps as detailed below. We retrieve the stage 1 products ({\tt rate.fits}) from the MAST archive and assign the world coordinate system (WCS) for each frame. Flat fielding is performed using the JWST calibration pipeline, with direct imaging flat reference files from the JWST Calibration Reference Data System (CRDS), as dedicated WFSS flat reference files are not currently available. The reduction uses CRDS version \texttt{11.17.15} and the context {\texttt{jwst\_1225.pmap}}. Our background subtraction process involves two steps: we first subtract a median background that is separately generated for each module and pupil. To further enhance data quality and eliminate any potential residual background, we employ additional subtraction using {\tt SExtractor} \citep{bertin1996}. To isolate the emission line
from the continuum, we follow the method described in \citet{Kashino23_grism}. This involves using a median filter technique that models the continuum with a sliding window, or  ``kernel". By subtracting this modeled continuum from the original data, we effectively isolate the emission line image.

With the fully calibrated emission line images, we are now prepared to extract the spectra of the sources of interest. This process includes assigning the sky coordinates of the selected galaxies and locating the 2D spectra along the dispersion direction using spectral tracing and grism dispersion models. To achieve accurate spectral traces, we measure the astrometric offset between the short wavelength (SW) direct images and the CLEAR catalog, applying this offset to both the direct images and the reduced grism images. This alignment ensures that the SW images and grism exposures are properly aligned with the CLEAR catalog. The spectral tracing, grism dispersion and sensitivity models are produced from multiple JWST/NIRCam Cycle 1 programs (PID: 1076/1536/1537/1538), the details are described in \citet{Sun23_grism}. We extract the 1D spectra using optimal extraction \citep{Horne86} and fit the 1D spectra with a single Gaussian profile to obtain the emission line flux.

\subsection{HST G102 and G141 Grisms}\label{sec:clear}
We utilize the archival HST grism observations from the CANDELS Lyman Alpha Emission At Reionization survey (CLEAR; GO-14227; PI: Casey Papovich), as detailed in \citet{simons23_clear}, which provided 12-orbit depth observations with the HST/WFC3 using the G102 grism, covering 12 fields across both the GOODS-N and GOODS-S. \citet{simons23_clear} expanded the original CLEAR survey dataset by incorporating observations from additional HST projects that utilize the HST imaging \citep{Giavalisco04, Grogin11, Koekemoer11, Momcheva16, Brammer12, Skelton14} and G102/G141 grisms \citep{Momcheva16, Pirzkal17} overlapping with the CLEAR footprint. The G102 observations come from programs GO-14227 (PI: C. Papovich),
GO-13420 (PI: G. Barro), and GO/DD-11359 (ERS; PI:
R. O’Connell). The G141 observations come programs GO-11600 (AGHAST;
PI: B. Weiner), GO-12461 (SN Colfax; PI: A. Reiss),
GO-13871 (PI: P. Oesch), GO/DD-11359 (ERS; PI: R.
O’Connell), GO-12099 (George, Primo; PI: A. Reiss), and
GO-12177 (3D-HST; PI: van Dokkum).

This approach achieved complete spectral coverage from 0.8 to 1.7 $\mu$m throughout the observed areas. Leveraging this rich dataset, \citet{simons23_clear} have compiled two spectroscopic catalogs that present emission line fluxes and redshifts, derived from a combination of photometry and grism spectroscopy. 
The specific observations can be accessed via \dataset[DOI: 10.17909/9cjs-wy94]{http://dx.doi.org/10.17909/9cjs-wy94}.

In our analysis, we use the redshifts and combined \ha\ + \nii\ flux from the CLEAR catalog. To correct for \nii\ contamination, we apply the \nii/\ha\ ratios as a function of stellar mass at \(z \sim 1.5\), as calibrated by the \(\text{KMOS}^{3D}\) program \citep{Wuyts_kmos}. For an in-depth understanding of the CLEAR survey's methodology, including survey design, data reduction processes, and the galaxy catalog compilation, we refer readers to \citet{simons23_clear}.



\begin{figure}
    \centering
    \includegraphics[width=0.47\textwidth]{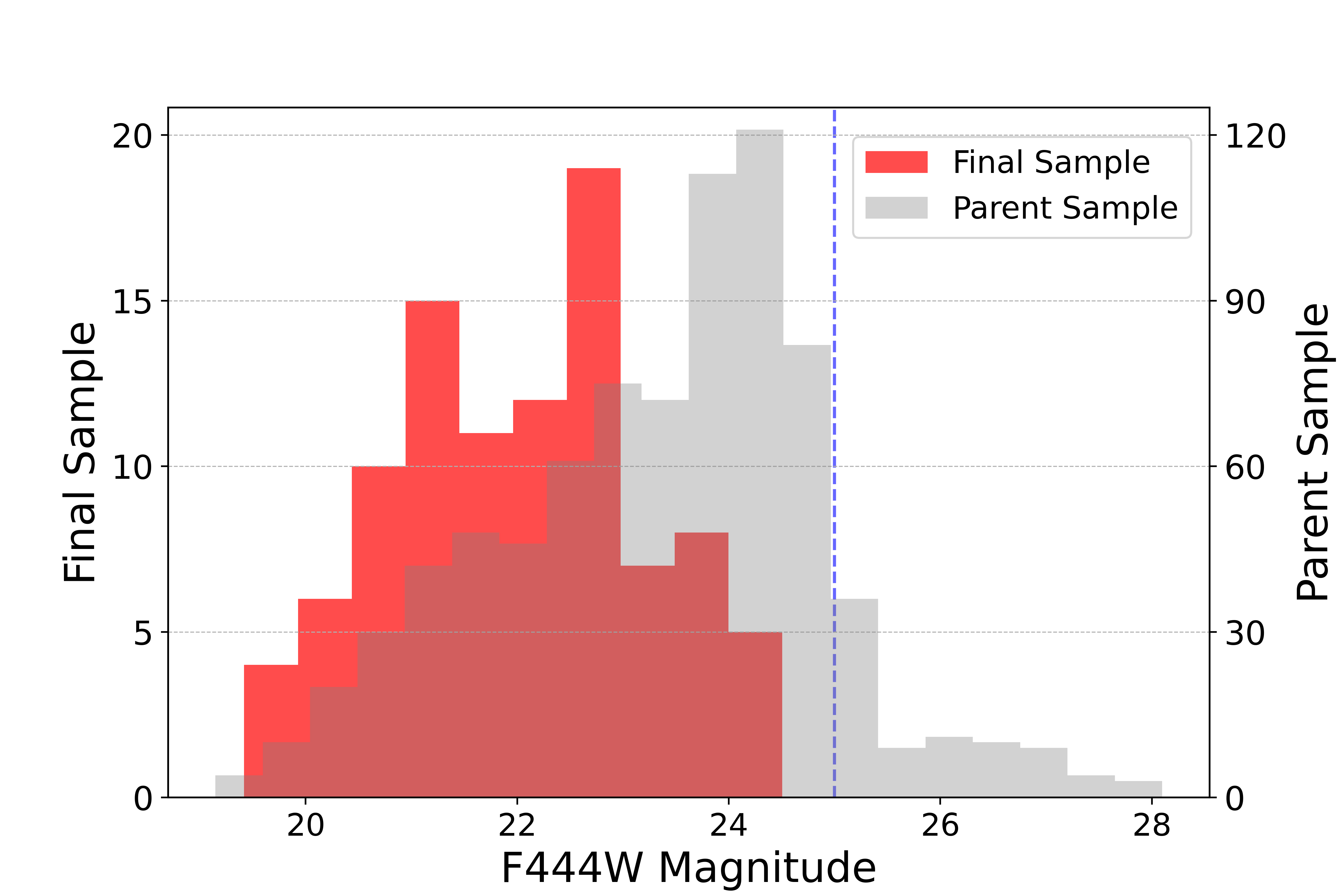} 
    \caption{Histogram showing the distribution of the F444W magnitude of our parent sample ({\it gray}) and final sample ({\it red}). The blue dashed line indicates our magnitude cut.}
    \label{fig:mag sample}
\end{figure}

\subsection{Sample Selection}\label{sec:sample selection}
We start with the NIRCam + HST catalog as described in Sec. \ref{sec:nircam catalog} and select sources for further analysis as follows:
\begin{itemize}

\item First, we perform a cross-match between the NIRCam+HST catalog (Sec.~\ref{sec:nircam catalog}) with the CLEAR spectroscopic catalog (Sec. \ref{sec:clear}) using a matching radius of $0.\!''5$, and retain only the galaxies with available spectroscopic redshift from the CLEAR survey. 
\item We further refine our selection to include only those galaxies within the redshift range of $1 < z < 1.6$, which aligns with the coverage of the NIRCam WFSS F444W filter for capturing the \pa\ line and the HST G141 grism for capturing the \ha\ line. We use the maximum likelihood redshifts (z$_{\textnormal{MAP}}$) for selecting galaxies at this step. The collection of galaxies that pass this phase of selection is referred to as our parent sample, which consists of 806 galaxies
\item  With the selected candidates, we check their overlap with the FRESCO footprint and obtain the WFSS spectra for galaxies within the WFSS coverage. We then perform visual inspections on the extracted spectra to remove galaxies with bright emission line or continuum contamination. Sources that are close to the WFSS spectra coverage boundary are also removed due to noise, especially at the outer part of the 2D spectra. Additionally, selected galaxies must have an AB magnitude brighter than 25 in F444W and both \ha\ and \pa\ detections with a SNR greater than 5.
\item Lastly, we conduct spectral energy distribution (SED) fitting for the selected galaxies in the previous step. To ensure reliable results, we further restrict our sample by including only those with the SED fitting results with a reduced $\chi^2_\nu$ value of less than 3. 
\end{itemize}
Our final sample consists of 42 galaxies from GOODS-S and 55 from GOODS-N. The distribution of F444W magnitudes for both our parent and final samples is displayed in Fig. \ref{fig:mag sample}.

\section{ANALYSIS} \label{sec:analysis}
\begin{figure*}
\plotone{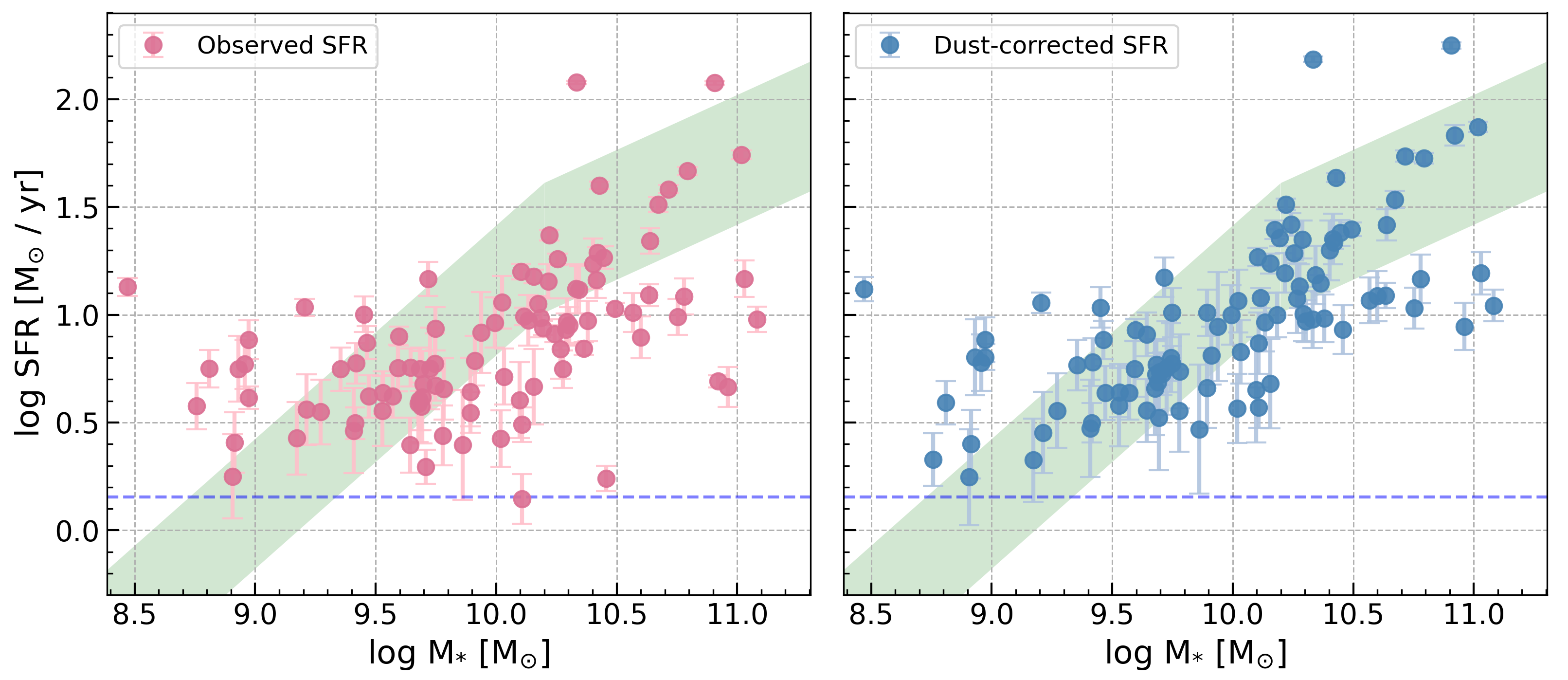}
\caption{({\it left}): Stellar mass versus SFR for our samples. Stellar mass is derived from SED fitting, SFR is calculated based on the \pa\ line flux without correction for dust extinction. Error bars for SFRs reflect flux uncertainties.The blue dashed line represents the SFR limit of 1.42 $\mathrm{M}_{\odot}\,\mathrm{yr}^{-1}$ at $z=1.5$, using a Pa$\alpha$ flux of $2 \times 10^{-18}\,\mathrm{erg\,s^{-1}\,cm^{-2}}$. The green-shaded region illustrates the $\pm0.3$ dex range of the star formation main sequence at $1<z<1.5$ derived by \citet{2014ApJ...795..104W}. ({\it right}): Analogous to the left panel, yet here SFRs are adjusted for dust extinction. Error bars incorporate uncertainties from both flux measurements and dust extinction A(\pa).
\label{fig:sfr}}
\end{figure*}

\label{sec:stellarmass_sed}
\subsection{SED Fitting}\label{subsec:sed}
We conduct SED fitting using the photometric data to estimate stellar masses. We use Code Investigating GALaxy Emission ({\tt{CIGALE}}, \citealt{2019A&A...622A.103B}). {\tt CIGALE} is particularly well-suited for analyzing data across a broad wavelength range, extending from X-ray to radio frequencies. Our SED fitting employs a delayed exponential model for star-formation history ({\tt sfhdelayed}), mathematically represented as ${\rm SFR}(t) \propto t \exp(-t/\tau)$. For the stellar population synthesis, we utilize the model from \cite{2003MNRAS.344.1000B}, assuming solar metallicity. The model incorporates dust attenuation based on the \cite{Calzetti2000} extinction law. A comprehensive description of the parameters used in our SED fitting can be found in \citet{liu23_smacs}. 

\subsection{\texorpdfstring{\pa\ and \ha\ Constraints on Nebular Reddening and SFRs}{Pa alpha and H alpha Constraints on Nebular Reddening and SFRs}}
\label{sec: dust extinction}
 In this work, we adopt Case B recombination assumption with ${\rm T_e = 10^{4}}$ K and ${\rm n_e = 10^{2}}$ ${\rm  cm^{-3}}$ as determined by photoionization models using CLOUDY version 17.02 \citep{Ferland17}. Under these conditions, the model predicts intrinsic intensity ratios of \ha/\hb\ = 2.79 and \pa/\hb\ = 0.305. By applying the \citet{Calzetti2000} reddening curve ($R_{\rm{v}}$ = 4.05) and comparing the observed \pa\ to \ha\ flux ratios with Case B predictions, we calculate \(E(B-V)_{\text{neb}}\) for our sample:
 
\[ E(B-V)_{neb} = \frac{2.5}{k(H\alpha) - k(Pa\alpha)} \log_{10}\left( \frac{Pa\alpha / H\alpha}{0.109} \right)\]

Here, $k(\lambda)$ represents the extinction coefficient at wavelength \( \lambda \), as prescribed by \citet{Calzetti2000}. From the calculated color excess, we derive the extinction values A(\ha) and A(\pa) using:

\[ A(\lambda) = k(\lambda) \cdot E(B-V) \]

Subsequently, with the estimated dust extinction from the \pa\ and \ha\ line flux ratios, we compute the intrinsic SFRs. The selection of the parent sample in this study relies on the \pa\ detection, which is sensitive to H{\footnotesize II} regions that signal active star formation. Thus, we expect the majority of our selected sources to be star-forming galaxies. To assess their star-forming activities more accurately, we derive dust-corrected SFRs using \pa\ line flux and color excess \(E(B-V)_{\text{neb}}\). The \pa\ flux is determined by fitting a single Gaussian profile to continuum-subtracted 1D spectra (Sec. \ref{sec:wfss}). Upon determining the intrinsic \pa\ flux using the \citet{Calzetti2000} approach, we convert it to the \ha\ flux based on the line ratio predicted under Case B recombination. SFRs are then estimated from the \ha\ luminosity following the SFR-$L_{\mathrm{H\alpha}}$ relation established by \citet{kennicutt98_2}:

\begin{equation}\label{eq1}
 \frac{\mathrm{SFR}}{\mathrm{M_{\odot}}\,\mathrm{yr^{-1}}} = 7.9 \times 10^{-42}\,\,\mathrm{L}_{\mathrm{H\alpha}}(\mathrm{erg\,s}^{-1})
\end{equation}

To align with the \citet{Chabrier03} IMF, we scale the SFRs by dividing them by a factor of 1.53 \citep{Driver13}. As illustrated in Fig.~\ref{fig:sfr}, most of our sample galaxies are located on or near the star formation main sequence predicted by \citet{2014ApJ...795..104W}. Our analysis indicates that the \pa\ line is a robust indicator of SFRs, relatively unaffected by dust extinction and offers a more direct measure of the ionizing radiation from young O and B stars. This is particularly true for less massive galaxies with a stellar mass below $\rm{10^{10}~M_\odot}$, which are less affected by dust extinction, as demonstrated in the comparison between the left and right panels of Fig.~\ref{fig:sfr}. 


\begin{figure*}
\centering
\includegraphics[width=0.95\textwidth]{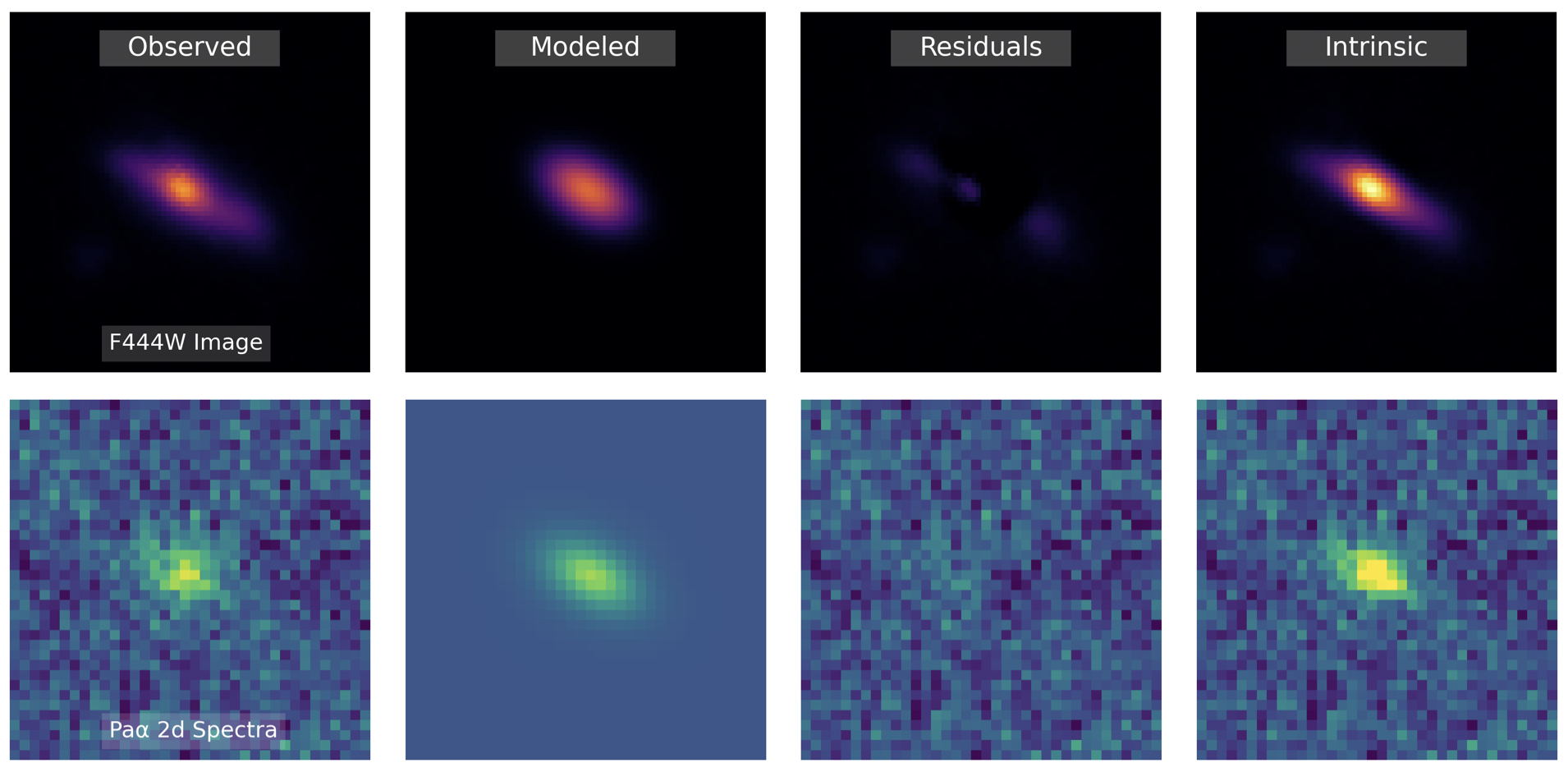}
\caption{Sequence from left to right illustrating intrinsic profile reconstruction: (1) The observed two-dimensional spectra. (2) The best-fit S\'ersic model convolved with the PSF. (3) Residuals derived from the difference between the observed spectra (1) and the PSF-convolved S\'ersic model (2). (4) The reconstructed intrinsic light distribution, obtained by adding the residuals to the best-fit S\'ersic model prior to PSF convolution. This reconstructed profile is used for subsequent galaxy size analysis.
\label{fig:fitting}}
\end{figure*}

\subsection{Size Measurements}\label{sec:size}
In our redshift range of interest, the F444W grism captures the \pa\ emission line, a robust tracer of SFRs, while the F444W image captures the rest-frame K-band image ($\sim$ 2 $\mu$m), providing an accurate measure of the galaxy's stellar mass unaffected by star formation history \citep[e.g.,][]{Kauffmann98, 2001ApJ...550..212B, Drory04}. Therefore, comparing the flux distributions from the F444W image and the F444W grism enables us to study the mechanisms governing star formation and mass assembly within galaxies, with minimal impact from dust extinction.

However, the NIRCam grism achieves a resolution of R $\sim$ 1600 at 3.95 $\mu$m, which corresponds to a velocity dispersion of $\sigma$ $\sim$ 80 km/s \citep{Nelson23, li23}. This resolution will lead to morphological distortion of emission line maps along the spectral axis of galaxies with a velocity dispersion greater than 80 km/s. As revealed by the \(\text{KMOS}^{3D}\) program \citep{Wisnioski15}, galaxies with a similar stellar mass range at z$\sim$1 exhibit rotation velocities ranging from 30 to 400 km/s, with the majority exceeding 150 km/s. Therefore, the 2D spectra we extract are not purely emission line maps; they also incorporate kinematic properties along the spectral axis. Overcoming such a degeneracy is possible with an additional medium-band filter that serves to extract the emission line map \citep[e.g.,][]{Nelson23}. Yet, this approach significantly narrows the applicable redshift range due to the coverage of medium filters. Recognizing these constraints, our analysis in this study is thus focused on examining the light profile distribution in the cross-dispersion direction, which remains unaffected by velocity dispersion. Our method involves the following steps:

\begin{itemize}

\item Generation of the Point Spread Function: We follow \citet{Matharu23} to incorporate PSF effects in our size measurements. Using \texttt{WebbPSF} \citep{2012SPIE.8442E..3DP, 2014SPIE.9143E..3XP}, we generate PSFs for both the 2D spectra and the continuum images. The F444W filter spans 3.9–5.0 $\mu$m, over which the PSF FWHM varies from $0.\!''12$ to $0.\!''16$. For the \pa\ 2D spectra --- taken at specific observed wavelengths corresponding to the galaxy redshifts --- we produce monochromatic PSFs based on each galaxy’s redshift. For the continuum images, we construct a wavelength-weighted PSF for F444W and apply it uniformly to all continuum data in subsequent steps.


\item Model Fitting: Following \citet{Szomoru10}, we use a S\'ersic $R^{1/n}$ \citep{1963BAAA....6...41S} to model both the galaxy images and the 2D spectra with \texttt{GALFIT} \citep{peng2002, peng2010}. First, we initialize the S\'ersic fitting using parameters derived from \texttt{SExtractor}. We then perform a second round of fitting, using the best-fit parameters from the initial fit, to obtain the final model.

\item Intrinsic profile reconstruction: Using the best-fit parameters determined above, we construct a S\'ersic profile and add back the residuals obtained from comparing the observed data with the PSF-convolved model. This profile, augmented by the residuals, is considered the intrinsic flux profile for subsequent size measurements. Although the residuals themselves are still influenced by the PSF, this technique has been shown to reliably recover the true flux distribution, even for galaxies that deviate from a pure S\'ersic profile \citep{Szomoru10}. Fig.~\ref{fig:fitting} outlines this reconstruction process schematically.

\item Flux Distribution Analysis: With the reconstructed intrinsic profiles for both 2D spectra and images, we compute the one-dimensional cross-dispersion light intensity profile by summing pixel values across each row. Specifically, we begin by creating cutouts of the 2D spectra and images, each with dimensions of $3.\!''0 \times 3.\!''0$; we then rotate F444W images to align with the 2D spectra, ensuring the horizontal direction is set as the dispersion direction. Additionally, we mask pixels that belong to other galaxies (based on segmentation maps) and those affected by artifacts (e.g., negative values from continuum removal), thus minimizing contamination from neighboring sources and bad pixels.

\item Size Measurement Along the Cross-Dispersion Axis: To compare the stellar and star-forming distributions within the same galaxy, we measure the half-light radius along the cross-dispersion direction of the processed images (PSF-deconvolved, orientation-aligned, and bad-pixel-masked). Because the grism data are shallower than the direct images, the SNR of the stellar component is typically higher than that of the star-forming regions. To ensure a fair comparison, we define a boundary using the location where the SNR of the 2D spectra drops to 1. This boundary is consistently applied to both F444W images and 2D spectra to account for the spectra’s reduced depth. We calculate the total flux within this boundary in each dataset. The half-light radius is then the distance along the y-axis (centered on the galaxy) at which the cumulative flux reaches half of the total.

\item Orientation Correction for \pa\ 2D Spectra: Because velocity-resolved 2D spectra are unsuitable for direct S\'ersic modeling, we measure their size using only the cross-dispersion axis. A single-axis measurement can introduce orientation-related uncertainties, especially for large or inclined galaxies. To mitigate this, we compute a correction factor: the ratio of (a) the S\'ersic-based radius from the F444W direct images (\texttt{GALFIT}) to (b) the half-light radius measured in the cross-dispersion direction of the \emph{same} F444W images. We then apply this ratio to the cross-dispersion half-light radius of the 2D spectra to account for a galaxy’s orientation.

\item Error Estimation: The uncertainty in the size measurement of the F444W images is provided directly by the \texttt{GALFIT} fitting, which outputs both the effective radius and its associated error. However, this uncertainty estimation cannot be directly applied to the \pa\ spectra. To estimate the uncertainty in the size measurement for the \pa\ spectra, we perform Monte Carlo simulations by generating 100 model galaxies. For each simulation, we randomly sample the S$\acute{\mathrm{e}}$rsic parameters from normal distributions centered on the best-fit values obtained from \texttt{GALFIT}, with standard deviations corresponding to their uncertainties. Using these sampled parameters, we create deconvolved S$\acute{\mathrm{e}}$rsic models and add the corresponding residuals from the original fit to account for deviations from the S$\acute{\mathrm{e}}$rsic model. We then measure the half-light radius for each of these modeled galaxies using the same method applied to the cross-dispersion direction. The final uncertainty is derived from the standard deviation of the half-light radii obtained from the simulations.

\item Summary: For F444W images, the half-light radius is taken directly from the \texttt{GALFIT}-derived S\'ersic fit. For 2D spectra, we first measure the half-light radius along the cross-dispersion axis of the PSF-removed 2D spectra, then multiply by the aforementioned ratio (from the matching F444W image) to correct for orientation. This approach enables a more reliable comparison between the stellar continuum size from the direct image and the \pa\ size from the 2D spectra, even though full S\'ersic fits are not feasible for \pa\ 2D spectra.

We provide the coordinates, redshifts, emission-line fluxes, dust attenuation estimates,
and size measurements for our sample. The contents of this catalog are summarized
in Table~\ref{tab:catalog}.

\begin{table}
\centering
\caption{Description of the Spectroscopic Catalog}
\label{tab:catalog}
\begin{tabular}{lll}
\hline
column & units & description \\
\hline
id &  & source identifier \\

ra & deg & right ascension (J2000) \\

dec & deg & declination (J2000) \\

redshift &  & JWST grism redshift\\

ha\_nii\_flux & erg s$^{-1}$ cm$^{-2}$ & H$\alpha$+[N\,II] emission-line flux \\

ha\_nii\_flux\_err & erg s$^{-1}$ cm$^{-2}$ &
uncertainty of line flux \\

paa\_flux & erg s$^{-1}$ cm$^{-2}$ & Pa$\alpha$ emission-line flux \\

paa\_flux\_err & erg s$^{-1}$ cm$^{-2}$ &
uncertainty of line flux \\

A\_ha & mag &
H$\alpha$ dust attenuation \\

A\_ha\_err & mag & uncertainty on $A_{\mathrm{H\alpha}}$ \\

A\_pa & mag & Pa$\alpha$ dust attenuation\\

A\_pa\_err & mag & uncertainty on $A_{\mathrm{Pa\alpha}}$
 \\

R\_F444W & kpc &
F444W half-light radius \\

R\_F444W\_err & kpc &
uncertainty on $R_{\mathrm{F444W}}$ \\

R\_paa & kpc &
Pa$\alpha$ half-light radius \\

R\_paa\_err & kpc &
uncertainty on $R_{\mathrm{Pa\alpha}}$ \\

\hline
\end{tabular}
\end{table}



\end{itemize}

%


\subsection{Stacking}

To enhance the SNR and measure the average properties of \pa\ and F444W profiles, we divide our sample into three stellar-mass bins (low, medium, and high), containing 32, 32, and 33 galaxies, respectively. Following the methodology described in \cite{nelson16}, we stack the \pa\ spectra and F444W images in each bin to create the mean profiles. Before proceeding with the stacking, we mask nearby sources using segmentation maps, deconvolve the PSF, and rotate the F444W images to ensure that the horizontal direction corresponds to the dispersion direction, as detailed in Sec. \ref{sec:size}. To prevent the final stacked profile from being dominated by brighter sources, we apply weights based on their F444W flux. According to \citet{nelson16}, deprojecting, rotating, or scaling the images does not significantly affect our main conclusions; therefore, we do not apply additional rotation to the dispersion axis-aligned profiles.

While \citet{nelson16} used a two dimensional S\'ersic fit, our data include velocity information along the dispersion direction, so we can only perform a one dimensional S\'ersic fit along the cross dispersion axis to measure the size of the stacked profile. To estimate the uncertainties in these 1D S\'ersic fits, we use a Monte Carlo approach: we generate 100 simulated profiles by adding random noise (drawn from stacked background) to the data. We then refit each noise-added profile with the same 1D S\'ersic method, and record the best-fit effective radius. Finally, we take the standard deviation of all 100 fitted radii as our estimate of the random uncertainty in the effective radius.

\section{Results} \label{sec:results}
In this section, we present the results of our analyses. First, we discuss the nebular reddening values derived from the \pa\ and \ha\ line ratio. Additionally, we present the dust extinction and SFRs inferred from \pa\ and \ha\ emission lines. We then explore differences in flux distribution between the F444W and \pa, and examine how size measurements correlate with stellar mass and sSFR to understand the structural build-up of galaxies. Finally, we discuss the average properties for stacked profiles of \pa\ and F444W.

\subsection{Dust Extinction}\label{sec: extinction}

\begin{figure}
    \centering
    \includegraphics[width=0.5\textwidth]{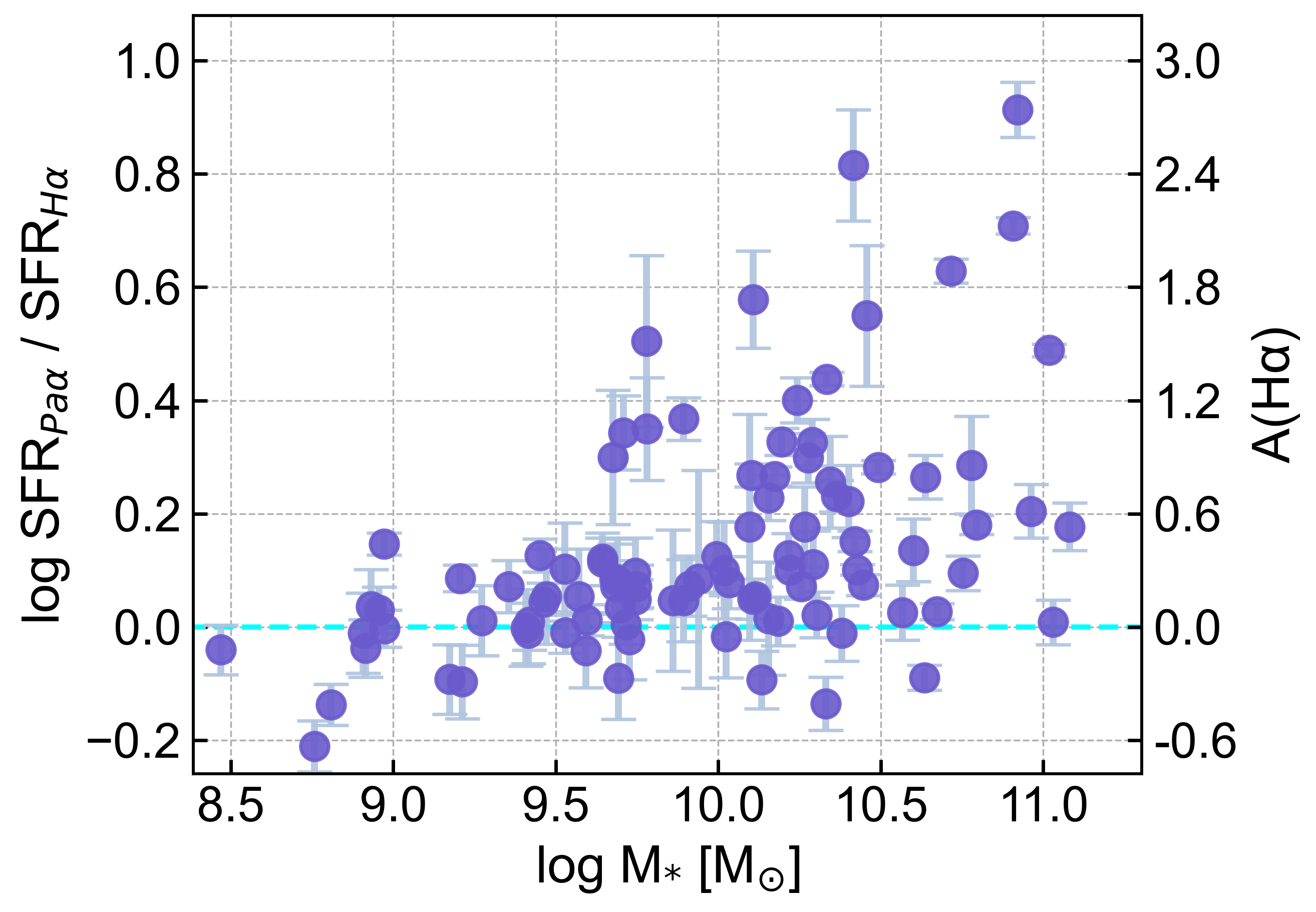}  
    \caption{Stellar mass versus ratio of observed SFR predicted using \pa\ and \ha\ respectively (left y-axis), and stellar mass versus A(\ha) (right y-axis). A(\ha) is calculated based on the observed ratio of \pa\ and \ha\ flux, and is thus proportional to the observed SFR ratio.}
    \label{fig:extinction}
\end{figure}

We examine the nebular reddening within our sample galaxies and investigate the relationship between dust extinction and stellar mass. In Sec. \ref{sec: dust extinction}, we derive the \(E(B-V)_{\text{neb}}\) using the ratio of the \pa\ and \ha\ emission lines. We then calculate the dust extinction A(\ha) and A(\pa) from these ratios. We present our findings in Fig. \ref{fig:extinction}, with error bars indicating the flux measurement uncertainties.

Our results indicate that the average dust extinction is higher in more massive galaxies, as inferred from the \pa-to-\ha\ line flux ratios. This is consistent with previous studies that employed Balmer-line ratios to assess dust extinction, as well as those using far-IR and UV continuum measurements \citep[e.g.,][]{Garn10, Zahid13, Reddy15, fudamoto20, Shapley23}. Based on the model predictions by \citet{Calzetti2000}, A(\pa) is significantly smaller than A(\ha) due to the much smaller k($\lambda$). Our results indicate that at our target redshift, A(\pa) is typically around $\sim$ 0.1 mag, with the median value increasing to only $\sim$ 0.2 mag in the most massive bin. While \pa\ is intrinsically fainter than \ha\ 
 (I$_{Pa\alpha}$/I$_{H\alpha}$ = 0.11 under Case B), this results demonstrate the robustness of \pa\ as a star formation indicator, less attenuated by dust extinction and more sensitive to the intrinsic SFR. Such conclusions are broadly in agreement with \citet{Reddy23}, who investigated the Paschen lines of 63 galaxies at redshifts $z = 1.0$--$3.1$ using NIRSpec. We note that a small fraction of galaxies in our sample exhibit a negative extinction. This happens when the observed ratio of \pa\ to \ha\ is less than the dust-free minimum value of 0.109, which is not physical. Such observations could be attributed to the relative faintness of the \pa\ line in comparison to the \ha\ line, possibly leading to incomplete detection of the galaxy's fainter regions in the 2D spectra. Furthermore, relying on the empirical calibration of the \nii/\ha\ ratio introduces additional uncertainties into our \ha\ flux measurements. Notably, recent studies with high signal-to-noise data have reported similar anomalies in hydrogen recombination line ratios, raising questions about the universal applicability of the Case B recombination assumption in extragalactic contexts. These complexities may involve scenarios such as optically thick nebulae or collisionally de-excited ionized gas under certain conditions \citep[e.g.,][]{Yanagisawa24, Scarlata24}.



\begin{figure*}
    \centering
    \includegraphics[width=\textwidth]{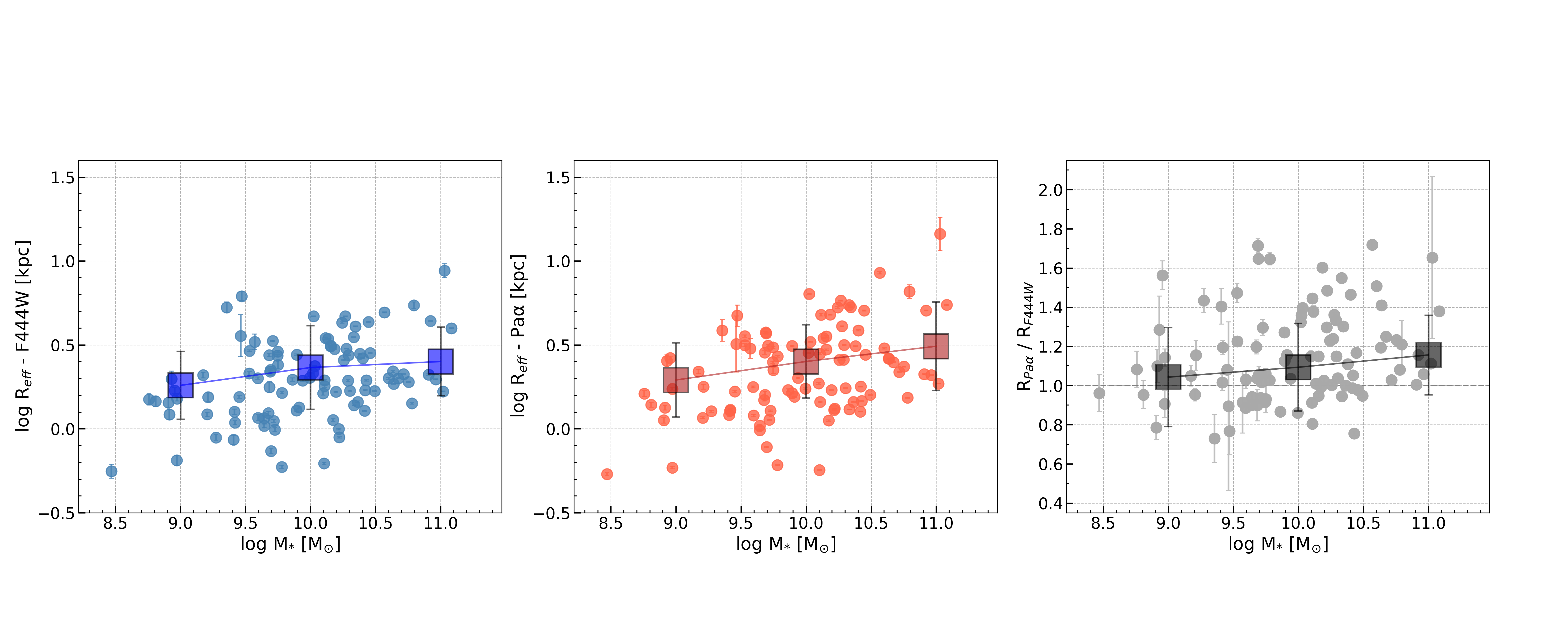} 
     \caption{({\it Left}): Effective radius from F444W images measured with \texttt{GALFIT} versus stellar mass. ({\it Middle}): Orientation-corrected effective radius from Pa$\alpha$ 2D spectra versus stellar mass. ({\it Right}): Size ratio between the two components as a function of stellar mass. The dashed line marks $\mathrm{R_{Pa\alpha}}/\mathrm{R_{F444W}} = 1$, indicating a consistent spatial distribution between star formation and stellar light. Large square symbols show the median values in three stellar mass bins (with 32, 32, and 33 galaxies in each bin, respectively). Error bars indicate the interquartile range, which captures the spread of the central 50\% of the data and reflects the intrinsic scatter within each bin.}
    \label{fig:mass_size}
\end{figure*}

\begin{figure*}
    \centering
    \includegraphics[width=\textwidth]{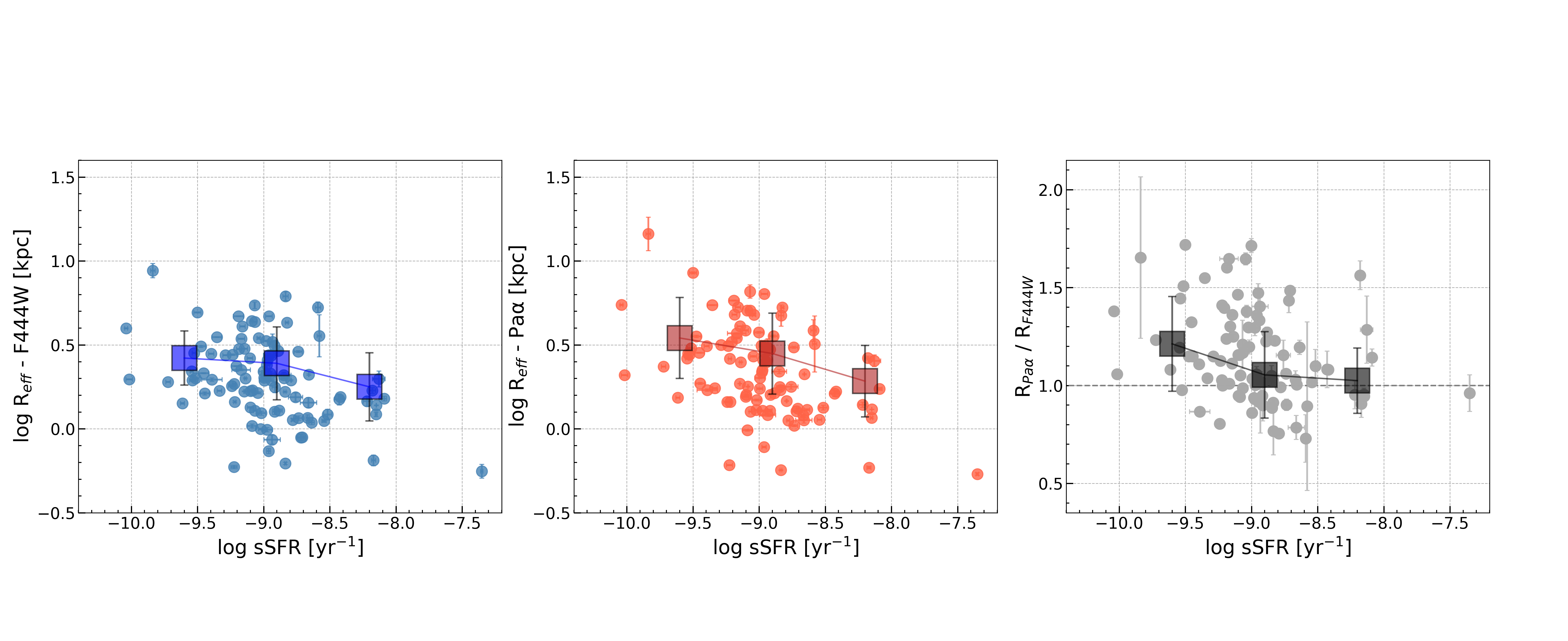} 
    \caption{Same as Fig.~\ref{fig:mass_size} but as a function of sSFR}
    \label{fig:ssfr_size}
\end{figure*}
\subsection{\texorpdfstring{\pa\ as a Star Formation Indicator}{Pa alpha as a Star Formation Indicator}}

\label{sec: resolved paa}

Thanks to the HST grism, previous studies \citep[e.g.,][]{nelson16, Vulcani16, Nelson19, Matharu22} have significantly advanced our understanding of star formation regions within galaxies through spatially resolved \ha\ emission line maps. As one of the most prominent hydrogen recombination lines, \ha\ is closely linked to ionized regions and serves as a reliable tracer of star formation. However, even \ha\ is susceptible to significant dust extinction. This presents a challenge for studying star formation locations as the dust amount may differ significantly across various regions of a galaxy \citep{nelson16_balmer, Tacchella18, Matharu23}. Such variability introduces notable uncertainties in quantifying galaxy sizes and analyzing morphological features. The utilization of spatially resolved \pa\ emission line map offers a solution to overcome these challenges. In this section, we present the galaxy sizes derived from \pa\ 2D spectra and F444W continuum data, offering insights into the spatial distributions of star formation activities and existing stellar mass within galaxies. In order to better understand the observed size difference in our galaxies, we investigate the relationship between the SFR-to-stellar size ratio, sSFR, and stellar mass.

\subsubsection{Half-light Radius of Individual Galaxies}
The size of galaxies serves as a fundamental metric for characterizing their structure, shedding light on bulge and disk growth and the properties of the encompassing dark matter halos \citep[e.g.,][]{Mo98, Kravtsov13, vanderWel14}. By comparing half-light radii measured for different components of galaxies, such as the stellar component and star formation component, we can gain vital information on how galaxies build their massive structures through vigorous star formation and how these activities vary across different evolutionary stages. Due to the limitation by the nature of our data, this study compares the half-light radius measured in the cross-dispersion direction. The calculation process for half-light radius is detailed in Sec. \ref{sec:size}. 

Furthermore, not only does galaxy size itself offer important clues about mass assembly, but its correlations with stellar masses, SFRs, and redshifts are equally crucial and have been extensively investigated. 
One of the key discoveries is the galaxy size-mass relation. Previous studies have shown that two main classes of galaxies, star-forming and quiescent, follow distinctly different size-mass relations \citep[e.g.,][]{Kauffmann03, Baldry06, vanderWel14}. One possible scenario to explain this discrepancy is that star-forming and quiescent galaxies accumulate their stellar populations through different processes \citep{vanDokkum15}. However, this is still a topic under much debate. To further investigate the origin, we study the correlation between size measurements and stellar mass to pinpoint where star formation activities occur within galaxies.

Fig.~\ref{fig:mass_size} shows how the half-light radius varies with stellar mass. The size of both \pa-traced star-forming disks and continuum-traced stellar components shows a weak increasing trend with stellar mass. However, it should be noted that discussing the mass–size relationship of \pa\ based on this `effective radius' alone may be misleading. This limitation arises because our \pa\ measurements are one-dimensional, taken along the cross-dispersion axis. Although we have corrected for galaxy orientation using information from F444W images, these corrections may not fully capture the intrinsic size of the \pa\ emission if the star-forming and stellar components have different spatial distributions.

Despite this limitation, our goal is to compare the flux distributions of the star formation and stellar components, which remains feasible with the current method. We find that less massive galaxies (i.e. galaxies with stellar mass $\rm{M_*<10^{9.5}\,M_\odot}$) exhibit similar sizes for both their SFR and stellar components. However, the ratio of these sizes tends to increase with stellar mass, albeit with considerable scatter. This result is consistent with the findings of \citet{nelson16}, who used HST grism observations to show that more massive galaxies often exhibit more extended \ha\ emission, while lower-mass galaxies have comparable sizes for their \ha\ and stellar continuum regions. Overall, these trends --- both here and in \citet{nelson16} --- support the inside-out growth scenario, suggesting that ongoing star formation occurs at larger radii relative to the already-established stellar distribution.

We further examine if the size ratio evolves with star formation per solar mass, specifically sSFR, as presented in Fig.~\ref{fig:ssfr_size}. For calculating sSFR, we adopt the dust-corrected SFRs as described in Sec. \ref{sec: dust extinction}, taking into account uncertainties from both flux measurements and dust extinction. Galaxies within the high sSFRs display compact structures in both their stellar and star formation components, with their size ratios indicating remarkably similar sizes.

\begin{figure}
    \centering
    \includegraphics[width=0.5\textwidth]{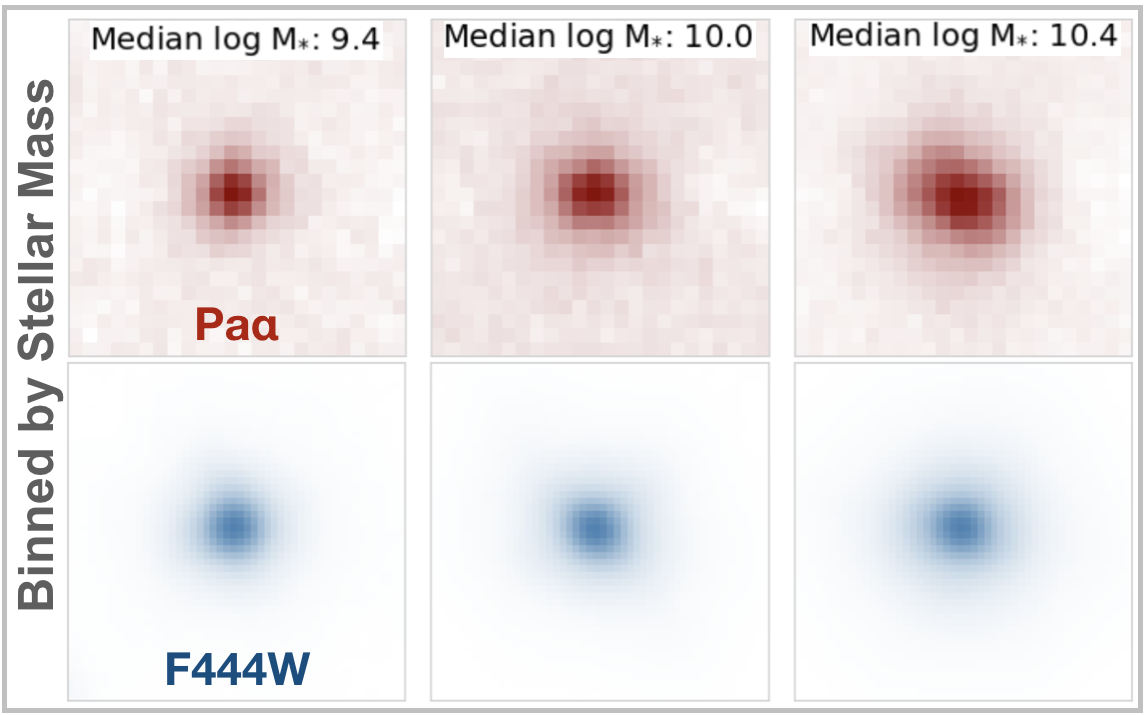}  
    \caption{Stacks of \pa\ ({\it red}) and F444W images ({\it blue}). From left to right, the sequence represents the stacked profiles of low, medium, and high stellar mass bins. The dimensions of the stacked images are \(1.\!''6 \times 1.\!''6\) for both \(\text{Pa}\alpha\) and F444W.}
    \label{fig:stack}
\end{figure}

\begin{figure*}
\plotone{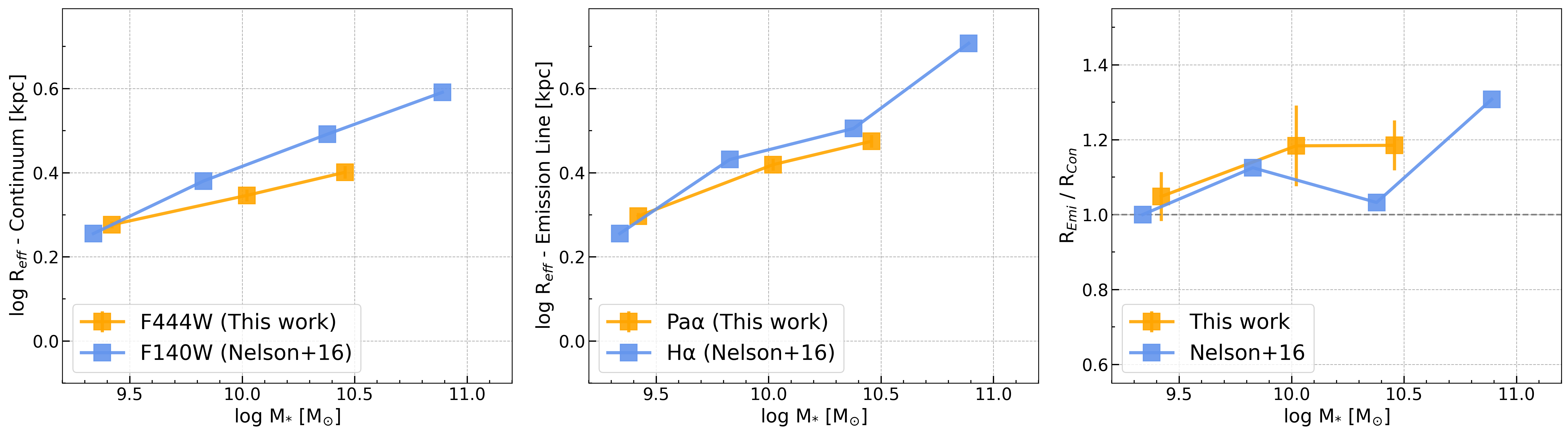}
\caption{({\it left}): Effective radius of stacked F444W images binned by stellar mass. ({\it middle}): Effective radius of stacked \pa\ 2D spectra binned by stellar mass. ({\it right}): Size ratio of two components versus stellar mass. In all three panels, the yellow lines indicate our results, while the blue lines show those from \citet{nelson16}.
\label{fig:stack_size}}
\end{figure*}
\subsubsection{Average Properties From Stacked Profiles}
As discussed in the Sec. \ref{sec: extinction}, the relatively faint nature of \pa\ emission line may lead to missing flux at the fainter part of \pa\ 2D spectra. To mitigate this issue, we categorize galaxies according to their stellar mass, and perform stacking analysis. Our stacked profiles for the three different bins of stellar mass are illustrated in Fig. \ref{fig:stack}. We subsequently measured the half-light radius of these stacked profiles in the cross-dispersion direction, with our results displayed in Figure \ref{fig:stack_size}. The stacked \pa\ profiles exhibit a more extended structure compared to the F444W images in the high and medium stellar mass bins, as evidenced by their larger effective radii. This trend is consistent with observations in individual galaxies, as shown in Fig. \ref{fig:mass_size}. 



\section{Discussions}
\label{sec:discussion}

\subsection{Dust, Star Formation, and Galaxy Mass Assembly}
\label{sec: galaxy mass assembly}
In Sec.~\ref{sec: resolved paa}, we investigated the half-light radius of \pa\ and F444W, and their potential correlation with stellar mass. 
Our main findings are illustrated in Fig. \ref{fig:mass_size} and \ref{fig:stack_size}, which show that galaxies, especially those with high stellar masses, exhibit \pa\ extent notably larger than their stellar continuum. These observations align with the findings of \cite{nelson16}, who reported a larger effective radius for \ha\ compared to F140W in a study of 3200 galaxies at $0.7<z<1.5$. Notably, the size difference was more pronounced with increasing stellar mass. While \cite{nelson16} observed a steeper size evolution with stellar mass than our results suggest, this difference could be attributed to the different tracers used, i.e.,
\ha\ and F140W in their study versus \pa\ and F444W in ours. Given that our tracers are expected to be more sensitive to attenuated emissions, the difference indicates that galaxy central regions of our galaxies systematically host more attenuated emissions. 

It should be noted that the half-light radius, determined from the integrated profile along the cross-dispersion axis, may not directly reflect the overall size of the galaxy due to the significant impacts of orientation and ellipticity. This factor might also contribute to the observed scatter in the relationship between size and stellar mass as shown in Fig. \ref{fig:mass_size}. Nevertheless, despite these limitations, the analysis of the ratio of galaxy sizes remains a robust metric. 

The observed increase in galaxy size ratio with stellar mass necessitates further investigation into the underlying physical processes that govern galaxy structural formation. It remains a challenging question to understand how galaxies form their central mass concentrations, particularly massive bulges, and evolve into mature, passive systems, given the presence of active star formation regions surrounded by dust. Our study provides a direct probe of star-forming regions, minimizing reliance on assumptions required for dust extinction corrections, such as UV slope and SED fitting techniques. Consequently, we anticipate that the regions traced by \pa\ will predominantly represent ionized hydrogen areas within the galaxies. Based on this foundation, and the positive correlation between \pa-to-stellar size ratio and increasing stellar mass (Fig. \ref{fig:mass_size}), our finding that most galaxies exhibit larger \pa\ sizes supports the inside-out growth model of galaxies \citep[e.g.,][]{vanDokkum10}. This model suggests that galaxies gradually build up their outer regions while the inner regions have completed their evolution more rapidly. 

Our study distinguishes itself from previous research that utilized \ha, which faces significant uncertainties due to  dust-obscured central regions. Our research confirms the scenario of extended star-forming regions, employing an independent tracer made feasible for the first time by NIRCam WFSS. Although \pa\ may still be significantly affected by dust extinction in extremely dusty starburst galaxies, as inferred from a recent study by \citet{Bik23} that studied a galaxy with A${_V}$ = 17.2, our sample primarily consists of normal star-forming galaxies, most of which have A(\pa) $<$ 0.2, with the dustiest having A(\pa) $\sim$ 0.6. 

\begin{figure}
\centering
\includegraphics[width=0.95\columnwidth]{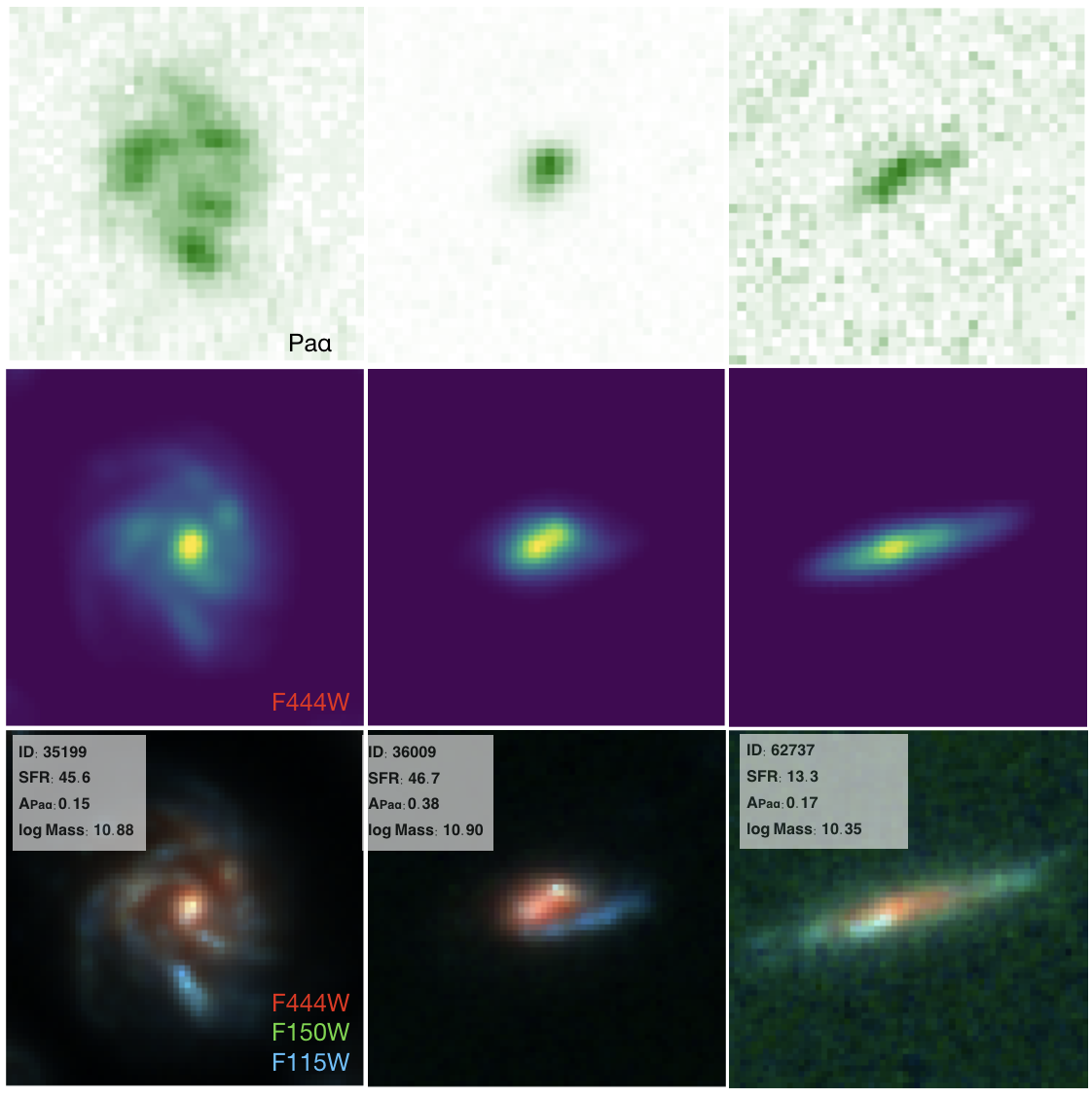} 
\caption{Cutout stamps showing \pa\ 2D spectra ({\it top}), F444W images ({\it middle}), and RGB images generated using F115W, F150W, and F444W images ({\it bottom}) for three galaxies in our sample. Each column corresponds to one galaxy, and each stamp measures $2.\!''5 \times 2.\!''5$. We also note the SFR ($\rm{M_\odot\,yr^{-1}}$), dust extinction at \pa\ wavelength (mag), and stellar mass (log $\rm{M_*/M_\odot}$) of each galaxy.}
\label{fig:massive}
\end{figure}

\subsection{Spatial Distribution of Star Formation in Massive Galaxies: Clumpy, Extended, and Compact} 
\label{sec: Massive Galaxies}
As discussed in Sec. \ref{sec: galaxy mass assembly}, we observe considerable scatter in the measurements of galaxy size ratios, particularly among medium-mass to massive galaxies (i.e. galaxies with stellar mass $\rm{M_*>10^{9.5}\,M_\odot}$). Within our sample, we identify three distinct flux distributions in the emission line maps of these galaxies: clumpy, compact, and extended. Even among galaxies with similar stellar masses and SFRs, their \pa\ flux profile can vary significantly. For example, as illustrated in Fig. \ref{fig:massive}, the left-hand galaxy exhibits clumpy star formation knots scattered throughout the disk, whereas the middle galaxy features a strongly centralized star-forming core, and the right-hand galaxy reveals radially extended star formation. 

Overall, stacked profiles in Fig. \ref{fig:stack_size} show that higher-mass galaxies tend to have larger star-forming regions; however, the individual scatter in the size ratio among these systems (right panel of Fig. \ref{fig:mass_size}) remains considerable, driven by variations in star formation patterns across different morphologies. Our findings align with previous work such as \citet{Swinbank12}, who reported a diverse set of \ha\ morphologies in nine star-forming galaxies at $0.8<z<2.2$, and \citet{Schreiber18}, who found that most of their 35 star-forming galaxies were disk-dominated, with a subset showing rings or clumps. The diversity observed in \ha\ is consistent with the variety of morphologies we have seen in \pa. In this section, we delve into the underlying physics behind the formation of clumpy, extended, and compact star formation regions in these galaxies. 

Both the clumps distributed at the outskirts of galaxies and the extended \pa\ can lead to a higher effective radius measured with \pa\ compared to that of the stellar continuum. Gaseous clumps, often contributing to the formation of bulges and disks, are regarded as key drivers of galaxy structure formation and evolution and are commonly observed in star-forming galaxies at $z>1$ \citep[e.g.,][]{Dekel2009, Dekel09}. Observations with HST have revealed that the presence of clumps --- and their properties such as color and location --- are tightly linked to the evolutionary and kinematic states of their host galaxies, and clumpy systems comprise a significant fraction of distant disk galaxies \citep[e.g.,][]{Schreiber11, Wuyts2012, Guo2015, Shibuya16, Guo2018}.

The existence of massive clumpy galaxies can be driven by dynamically unstable gas, which leads to the formation of large star-forming clumps within the disk \citep{Krumholz18}. Alternatively, it may reflect gas-rich minor mergers that directly introduce clumps or trigger their formation \citep{Agertz09, Mandelker14}. In either scenario, clumps can significantly affect both the observed radial distribution of star formation and the overall structure of their host galaxies.

To further understand the formation mechanisms of these complex structures, observations of both ionized and molecular gas are essential. There are only a handful of studies focusing on clumpy galaxies; recently, \citet{Bik23} analyzed the dusty, star-forming galaxy GN20 at $z\sim4$ using MIRI/MRS, detecting spatially resolved \pa. Their emission line map revealed that \pa\ emissions are concentrated in four distinct clumps. Additionally, GN20 was observed with ALMA \citep{Hodge12, hodge16}, where the CO(2-1) emission also revealed a similar clumpy structure. These findings suggest that the galaxy is in the late stages of a major merger, implying that the clumps in the gas-rich disk are a result of this merger, while the central starburst is directly driven by the event. Similarly, \citet{Cochrane2021} reported bright, extended structures in a $z=2.2$ galaxy, with clear clumps in \ha\ and extended dust continuum emissions. This disordered morphology and extreme star formation are linked to a merger event. Although our sample primarily consists of main sequence galaxies, it is possible that some of the clumpy structures we observed are triggered by merger activities. Such sub-structures remain an active area of exploration, and our study indicates that clumpy star-forming populations can be efficiently studied with WFSS. To fully comprehend the impact of dust and molecular gas in shaping the characteristics of H{\footnotesize II} regions, further investigations using sub-mm observations are crucial \citep[e.g.,][]{Liu25}. 

The extended star formation region observed in the right panel of Fig. \ref{fig:massive} has been well-studied in previous \ha\ observations utilizing HST grism and adaptive optics \citep[e.g.,][]{nelson16,nelson16_balmer, Vulcani16, Matharu22}. Unlike rest-frame UV-optical light, \ha\ emission suffers less attenuation by dust and is closely linked to the star formation activities of massive stars. Given the relatively short lifespans of massive stars, their observed population offers a direct measure of the current SFR \citep[e.g.,][]{kennicutt98_2, Altenburg07}. It was reported that when galaxies assemble their massive structures, star formation typically starts in the central region and then spreads outward \citep{vanDokkum10, morishita15, nelson16}. This process helps build up the galaxy’s outskirts, leading to the larger size of star formation regions observed in massive galaxies.

Compact star formation presents another intriguing scenario for further investigation. One hypothesis for the most massive galaxy populations is that they undergo a process known as ``compaction", which leads to the evolution of massive bulges over cosmic time \citep{dekel14,Zolotov15, Lacerda20, Marques22}. Subsequent starbursts and/or AGN feedback may then drive these galaxies toward quiescence \citep{genzel14, Yesuf14}, resulting in quenched, dense cores \citep{Damjanov09, vanDokkum09, szomoru12, morishita&ichikawa16}. Recent studies have elucidated the formation mechanisms of compact galaxies. \citet{Franco20} highlighted the significant role of compact galaxies at z $\sim$ 2.5 - 3, associating them with short depletion timescales. Simulations such as Illustris suggest that these galaxies often form early in extreme environments characterized by an abundance of gas and/or in (post-)merging systems where large inflow of gas trigger intense starbursts \citep{Wellons15}. The acquisition of external gas is $\mathbf{believed}$ to be responsible for initiating secondary star formation within the core regions of these galaxies that result in their compact morphology \citep{Chen16}. Importantly, such populations might be underrepresented or even missed in previous \ha\ analyses, at redshifts $z = 1-2$, our target epoch corresponding to the peak of cosmic star formation activity, dust extinction affecting \ha\ can vary significantly, with magnitudes ranging from $\sim1$, to $>3$ \citep[e.g.,][]{Schreiber09, sobral12, Kashino13, Koyama19}. Moreover, dust extinction can vary widely across a galaxy. As discussed in \cite{Matharu23}, dust attenuation is not a constant value within galaxies; instead, the spatial distribution of dust attenuation can vary across different stellar masses. A recent study by \cite{Cochrane2021} investigated a massive galaxy at $z = 2.2$ using multi-wavelength data and found A(\ha) values ranging from $\sim 2-3$ in the outskirts to $>$ 5 in the central region. Interestingly, their \ha\ emission line map, obtained using SINFONI/VLT, showed a clear offset from the peak of the dust continuum emission traced by ALMA. \cite{tadaki14} also reported color gradients in galactic clumps, with central clumps being redder, indicating that dusty star-formation activity is concentrated in these nuclear regions. Such dust gradients introduce significant complexity to the analysis of \ha-traced H{\footnotesize II} regions, potentially biasing detections toward galaxies with extended star formation regions. Galaxies featuring compact, yet dusty, star formation regions might remain undetected in \ha\ observations. Therefore, compact star-forming galaxies might still be understudied in previous \ha-based work. Future surveys using better tracers less affected by dust, such as \pa, will be crucial to provide a comprehensive view of the different paths of galaxy structure formation.

\section{Summary and Future Prospect}
\label{sec:summary}
This paper presented a comprehensive analysis of resolved \pa\ emissions within 97 galaxies at $1<z<1.6$ in the GOODS-North and GOODS-South fields utilizing JWST NIRCam WFSS. By combining this new dataset with the previous legacy HST grism observations, we were able to not only secure the redshift determination but also robustly determine the dust extinction for our sample galaxies. The main findings of our research are summarized below:
\begin{itemize}
    \item We calculated the dust extinction values, A(\ha) and A(\pa), for our sample of 97 galaxies and confirmed that \pa\ is an excellent tracer for star formation activities in H{\sc ii} regions, as it is not significantly attenuated by dust. We observed a maximum A(\pa) of less than 0.6, and noted that low-mass galaxies are almost attenuation-free. This establishes a solid basis for analyzing spatially resolved star-forming activities within galaxies with minimal impact from dust.
    
    \item We utilized HST and JWST photometric data to estimate stellar masses through SED fitting and derived SFRs from \pa\ emission line flux. Our sample predominantly features main sequence galaxies, and the \pa-derived SFRs exhibit minimal variation before and after dust extinction correction, with the exception of a few of the most massive galaxies.
    \item 
    From the analysis of individual and stacked emission maps of \pa\ and F444W, we found that for galaxies with medium and high stellar mass, the \pa\ show a larger size compared to the stellar size, whereas for galaxies with low stellar mass, the dimensions remain comparable. 

    \item We explored the size evolution of stellar and star-forming regions, and their ratios, across different stellar mass bins. In low-mass galaxies, star formation is concentrated centrally, resulting in compact sizes with stellar and star-forming regions being similar in dimension. In contrast, medium- and high-mass galaxies show more diverse star formation regions that can be extended, clumpy, or compact. This diversity indicates varying star formation mechanisms, potentially driven by external gas inflow or mergers, which become more prominent as stellar mass increases.
\end{itemize}


\section*{Acknowledgements}
We sincerely thank the anonymous reviewer for the constructive comments, which have significantly improved the quality of this manuscript. We acknowledge the teams of the JWST observation programs \#1180, \#1895, and the HST observation program \#14227 for their hard work in designing and planning these programs, and for generously making their data publicly available. The data presented in this paper were retrieved from the Mikulski Archive for Space Telescopes (MAST) at the Space Telescope Science Institute, the specific observations analyzed can be accessed via \dataset[DOI: 10.17909/z0sb-mk09]{http://dx.doi.org/10.17909/z0sb-mk09}.  We thank Fengwu Sun for helping with the data reduction. We thank Jasleen Matharu, Yongda Zhu and Yunjing Wu for the valuable discussion. 
ZL is supported by the Japan Society for the Promotion of Science (JSPS) through KAKENHI Grant No. 24KJ0394 and by JST SPRING, Grant No. JPMJSP2114. This work is also supported by JSPS KAKENHI Grants No. 24H00002 (Specially Promoted Research by T. Kodama et al.) and No. 22K21349 (International Leading Research by S. Miyazaki et al.). Support for this study was provided by NASA through grants HST-GO-15804 and HST-GO-17231 from the Space Telescope Science Institute, which is operated by the Association of Universities for Research in Astronomy, Inc., under NASA contract NAS 5-26555.

\appendix
\label{appendix}

\section{Flux and Extinction Measurements}

\begin{figure}
\centering
\includegraphics[width=0.9\textwidth]{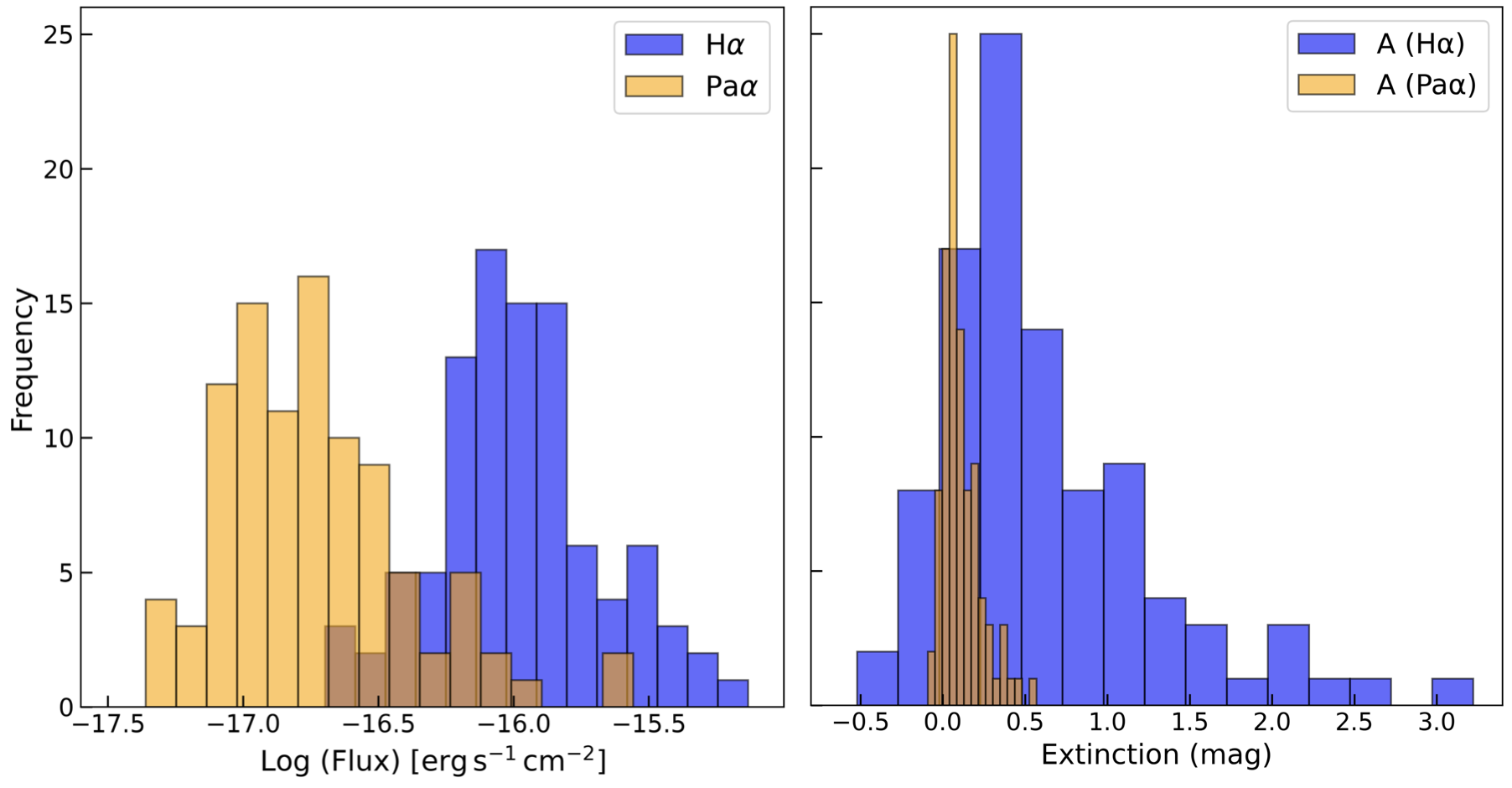} 
\caption{Histogram showing the distribution of the flux measurements ({\it left}) and the extinction at the corresponding wavelengths ({\it right}) for \ha\ and \pa, respectively. The \ha\ flux values have been corrected to remove the \nii\ contribution, as detailed in Sec. \ref{sec:clear}.}
\label{fig:measurement}
\end{figure}

{Fig.~\ref{fig:measurement} shows the distributions of emission line flux and extinction at the corresponding emission line wavelength for our sample of 97 galaxies. The \ha\ emission line flux has been corrected for the \nii\ contribution based on the stellar mass calibration, as detailed in Sec.~\ref{sec:clear}. 

Dust extinction at the emission line wavelength is calculated by comparing the observed \pa/\ha\ flux ratio with the Case B prediction and then using the \citet{Calzetti2000} law (assuming $R_{\mathrm{v}} = 4.05$) to determine the extinction value. We adopt this method because, as shown by \citet{Reddy23}, the \citet{Calzetti2000} extinction law remains valid even at \pa\ wavelengths.}

\section{Robustness of Size Measurements Across Imaging Depths}
\begin{figure}
\centering
\includegraphics[width=0.5\textwidth]{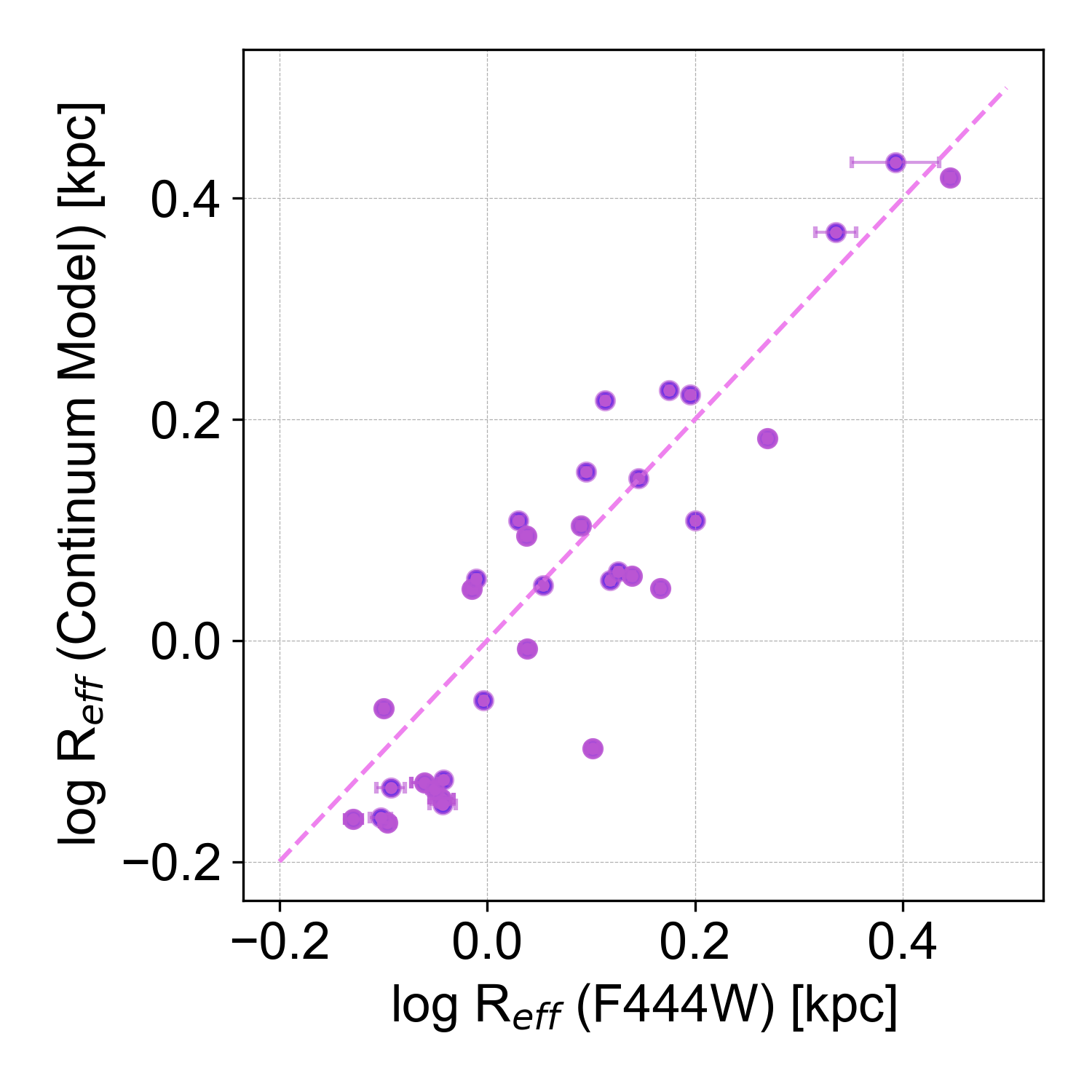} 
\caption{Comparative distribution of half-light radii derived from F444W images (x-axis) and modeled continuum (y-axis), both in the cross-dispersion direction.}
\label{fig:continuum}
\end{figure}

{One might question our size measurements due to the differing depths of the F444W imaging and WFSS spectroscopy, with the former being significantly deeper. Such a depth discrepancy could result in an underestimation of the size of star formation regions, as the fainter outskirts of galaxies may remain undetected. Ideally, a comparison between the continuum captured by WFSS and the observed emission lines would help to mitigate these discrepancies. However, practical implementation of this method faces significant challenges as the continuum captured in dispersed spectra is frequently contaminated by nearby continua, thereby complicating the accuracy of measurements. This issue is particularly pronounced for larger galaxies, where the continuum is more likely to overlap with other sources. Therefore, our analysis relies on direct images from F444W. To further assess the potential effects of this depth difference, we carefully inspect each source and select those without significant contamination from nearby galaxies. We measure their sizes using the modeled continuum (see Sec. \ref{sec:wfss} for the continuum modeling process), applying the same measurement method as used for F444W images and 2D spectra. The results, illustrated in Fig. \ref{fig:continuum}, show that measurements based on the modeled continuum and F444W are generally consistent. Therefore, we assert that the depth difference does not significantly impact our conclusions. Even if the sizes of the 2D spectra are underestimated, our findings that the star formation regions traced by 2D spectra are more extended remain unaffected. }

The inherent complexities in galaxy morphology and kinematics, particularly the elongation of morphologies along the spectral axis due to kinematic distortions, prevent us from obtaining accurate and separate information on spatial and spectral information. As discussed in \citet{Nelson23}, for high EW emission lines, such as \ha, a combination of medium-band filters and WFSS can provide insights into both the velocity and morphology of the emission line maps. However, this approach may be challenging for weak lines or for fields without medium-band filters on the targeted line. To apply this method to our research objectives, specifically using \pa\ to trace dust-free star formation, a larger sample size would be required. 
A future joint analysis involving multiple programs with available medium-band and WFSS observations will offer more opportunities to study galaxies with strong \pa\ emission lines.

\software{\texttt{Astropy} \citep{Astropy13,Astropy18},
          \texttt{SExtractor} \citep{bertin1996},
          \texttt{CIGALE} \citep{2019A&A...622A.103B},
          \texttt{GALFIT} \citep{peng2002, peng2010}}

\bibliography{sample631}{}
\bibliographystyle{aasjournal}



\end{document}